\documentstyle[preprint,aps]{revtex}
\begin{document}
\preprint{TIFR-TH/98-07, MIT-CTP-2723,}
\title{Hawking Radiation from Four-dimensional Schwarzschild
Black Holes in M-theory}
\author{Sumit R. Das$^1$, Samir D. Mathur$^2$ and P. Ramadevi$^1$}
\address{$^1$ Tata Institute of Fundamantal Research, Homi Bhabha Road, 
Mumbai 400 005, INDIA}
\address{$^2$ Center for Theoretical Physics, 
Massachusetts Institute Of Technology, Cambridge, MA 02139, USA}
\maketitle
\begin{abstract}

Recently a method has been developed for  relating four dimensional
Schwarzschild black holes in M-theory to near-extremal black holes in
string theory with four charges, using suitably defined ``boosts'' and
T-dualities. We show that this method can be extended  to obtain the
emission rate of low energy massless scalars for the four
dimensional Schwarzschild hole from the microscopic picture of radiation
from the near extremal hole.

\end{abstract} 
\pacs{PACS\,  04.70.Dy, 11.30.Pb, 11.25.-w } 
\vskip2pc]

\def\vk{{\vec k}}
\def\ben{\begin{equation}}
\def\een{\end{equation}}
\def\bea{\begin{eqnarray}}
\def\eea{\end{eqnarray}}
\def\bean{\begin{eqnarray*}}
\def\eean{\end{eqnarray*}}

In the past couple of years there has been considerable progress in
understanding the statistical basis of the thermodynamics of
near-extremal five and four dimensional black holes in string theory
\cite{reviews}. Further, a simple effective model gives the correct
low energy Hawking radiation from such holes. A natural question to
ask is: Can we get the emission from neutral (i.e. Schwarzschild)
holes as well?

Recently there has been some progress in understanding Schwarzschild
black holes in M theory.  The basic idea is to use 11-dimensional
Lorentz invariance properties to relate Schwarzschild holes boosted
along $x^{11}$ to string theory states carrying Ramond-Ramond charges
\cite{bfks}. In \cite{dmkr} a concrete map was found which relates
Schwarzschild strings and black $p$-branes. It was found that the
precise relationship is not through a genuine boost in the compact
direction, but through a boost in the covering space. The boosted
coordinate is then re-compactfied on a circle of radius which is
related to the original radius by Lorentz contraction. This is not an
exact symmetry of the theory, but provides a concrete map at the classical
level. T-dualities can generate other charges from the momentum
charge, and a combination of boosts and T-dualities may be used to
relate Schwarzschild holes with other known black holes in string
theory carrying charges, as in \cite{tseyta}. Specific maps which
relate a five-dimensional Schwarzschild black hole with the standard
five dimensional black hole in string theory with three large charges
were given in \cite{dmkr}. Using similar steps one can map the four
dimensional Schwarzschild hole to three sets of near-extremal 5-branes
of M-theory, intersecting in a common line along $x^{11}$, and
carrying momentum along this direction \cite{aeh}.  This is a
description of a four dimensional black hole with four charges in M
theory \cite{cvev} The entropy of the latter near extremal
hole is known  to follow from a microscopic calculation \cite{fourd,mbrane}, 
and thus we get the entropy of the neutral hole \cite{aeh}.

In the microscopic picture we  can also get the emission from the above 
near extremal model
\cite{gk} , so one wonders if the maps allow us to predict the
emission from neutral holes. We show in this paper that, using a
general relation between absorption cross-sections derived in
\cite{dmkr}, such is indeed the case, at leading order in the energy
if we assume during the calculation that 
the radius of $x^{11}$ is large enough.  The important issue turns
out to be the way the maps act on the emitted quantum while they
transform the black hole - it turns out that the low energy scalars
emitted from the neutral hole indeed map to quanta whose emission we
can compute from the microscopics of the near extremal hole.

We will consider 11-dimensional M-theory compactified on a space
$T^p \times S^1$. The torus has sides $L_i, i=1,\cdots 6$ and the circle, which
will be taken to be along the $x^{11}$ direction, has a radius $R$. 
In this space a  ``Schwarzschild string'' is a product of
a Schwarzschild black hole in the noncompact $(10-p)$ dimensions and
flat space along the $(p+1)$ compact dimensions. (This is thus a
$(p+1)$ black brane. We call this a string since it extends along $x^{11}$.)
The 11-dimensional metric is
\bea
\label{eq:one}
ds_{11}^2 & = & -(1-({r_0 \over r})^n)dt^2 
+ {dr^2 \over (1-({r_0 \over r})^n)} \\ \nonumber
& &  + r^2 d\Omega_{n+1} + dz^2
+ \sum_{i=1}^p (dx^i)^2,
\eea
where
$n = 7-p$ and $r^2 = \sum_{i=p+1}^9 (x^i)^2$.
When the Schwarzschild radius is smaller
than the radius $R$, a second object - the periodic Schwarzschild 
black hole - which is a periodic version of a $(11-p)$ dimensional
Schwarzschild hole with $x^{11}$ as one of the transverse directions -
is entropically favorable.
This is the subject of \cite{bfks},\cite{haylo}-\cite{bfk}.
A proposal for the microscopic description from this phase has been given
  in terms 
of a gas of D0 branes \cite{horomart,bfk}. 

The map which relates the Schwarzschild string to a charged hole consists
of a boost in the covering space in which $x^{11}$ is noncompact, but the
other $p$ directions are still compact
\ben
z' = z\cosh\alpha + t\sinh\alpha,~~~~
t' = t\cosh\alpha + z\sinh\alpha.
\label{eq:ktwo}
\een
The boosted coordinate $z'$ 
has to be then compactified on a radius $R'$ which
is related to $R$ by a standard Lorentz contraction
\ben
R' = {R / \cosh \alpha}.
\label{eq:ksix}
\een
By standard Kaluza-Klein reduction along $z'$ (as in \cite{horne})
the resulting metric
then represents a RR charged black hole in $(10-p)$ dimensions.
Applying T-duality along the $p$ compact directions results in
a black $p$-brane.
The relationship (\ref{eq:ksix}) ensures that the energy and momenta
transform correctly and the semiclassical entropy is kept invariant
under the boosts \cite{dmkr}. For further application of these ideas to seven
dimensional Schwarzschild holes see \cite{er}.

Furthermore it was shown in \cite{dmkr} that not only the entropy but the
absorption cross-sections (and hence Hawking radiation rates) of
neutral and charged black holes may be related using these ``boosts''
at least when $R$ and $R'$ are large enough to allow us to ignore the
quantization of momentum in this direction.  If $\sigma (\omega,q,\vk;A)$
denotes the absorption cross-section of some particle of energy
$\omega$, momentum $q$ along $x^{11}$, and transverse momentum
$\vk$ by the black hole ${\cal A}$, then the absorption cross-section
$\sigma'(\omega',q',\vk';{\cal A}')$ of the transformed particle with
energy-momentum given by $(\omega',q',\vk)$ by the transformed black hole
${\cal A}'$ is given by
\bea
& & \sigma'(\omega',q',\vk';{\cal A}') = {\omega \over \omega'}~
\sigma (\omega, q, \vk ; {\cal A}),  \nonumber \\
& & q'= q \cosh \alpha + \omega \sinh \alpha, ~~~
\omega' = q \sinh \alpha + \omega \cosh \alpha.
\label{eq:zthree}
\eea
The emission rate $\Gamma (k)$ is related to $\sigma$ by
\ben
\Gamma(\omega,q,\vk) = {\sigma (\omega,q,\vk)
\over e^\xi \mp 1}~{d^d \vk \over (2\pi)^d};~~~\xi = {(\omega - q \phi) / T},
\label{eq:decay}
\een
where $T$ is the temperature of the black hole, and
$\phi$ is the potential
of the Kaluza-Klein gauge field at the horizon (in 10d language) and $d$
is the number of transverse dimensions. 
The relation (\ref{eq:zthree}) does not depend on {\it how} the cross-section
is calculated, but follows from the fact that decay rates decrease by
a time dilation factor.
Hence if one had a microscopic derivation of the
cross-section for the charged black hole one may use this to obtain a
microscopic derivation of the cross-section for the neutral black
hole. Relations similar to above have been used to predict the
{\em total} emission rates in the D0 gas model in \cite{bfk}.

We now apply the above formula for four dimensional black holes.
Consider a four dimensional Schwarzschild black hole with area $A_0 =
4\pi r_0^2$ and temperature $T_0 = 1/4\pi r_0$ tensored with $T^7$
(the metric is given by (\ref{eq:one}) with $p = 6$) emitting some
massless scalar, e.g.  a longitudinal component of the metric $h_{12}$
with some momentum $k_0$ along a transverse direction, say $x^7$. The
energy of this particle is thus $\omega_0 = |k_0|$. The universal low
energy absorption cross-section is given by the area of the two 
dimensional horizon
\cite{absorption} 
\ben 
\sigma_0 = A_0,
\label{eq:zsixa}
\een
while the parameter appearing in the thermal factor is $\xi_0 = \omega_0/T_0$

Now perform the following operations.

\begin{enumerate}

\item Boost along $x^{11}$ by parameter $\alpha_1$

\item T-dualize along $(1234)$

\item Boost along $x^{11}$ by parameter $\alpha_2$

\item T-dualize along $(1256)$ 

\item Boost along $x^{11}$ by parameter $\alpha_3$

\item T-dualize along $(1234)$

\item Boost along $x^{11}$ by parameter $\alpha_4$

\item T-dualize along $(1256)$

\end{enumerate}

The entropy is the same at all stages.
At every step we will denote the string coupling by $g_n$, the string
length by $l_n$, the radii of the torus by $L_i^{(n)}$
and the $x^{11}$ radius by $R_n$ where $n = 1 \cdots 6$.
The first seven steps were used in \cite{aeh}. For us, however, the last
step is important.

With every boost by a parameter $\alpha$ the two dimensional
horizon area  and the temperature (in Planck units)  changes as
\ben
A \rightarrow A' = A \cosh \alpha,~~~~~~
T \rightarrow T' = {T / \cosh \alpha}.
\label{eq:zfive}
\een
The absorption cross-sections are
related by (\ref{eq:zthree}). A T-duality keeps
the cross-section invariant, but changes the nature of the black hole
as well as that of the emitted particle.  Our strategy will be to
first obtain a prediction for the semiclassical absorption
cross-section at the final stage, starting from the known semiclassical
answer for the absorption cross-section at the initial stage. Finally
the former will be compared with a microscopic calculation performed
at the last stage.

After the first step one has $R_1 = R/\cosh \alpha_1$. 
The black hole has $x^{11}$ momentum which
is a 0-brane charge in ten dimensional language. Its area $A_1$ and
temperature $T_1$ and rank-1 potential at the horizon $\phi_1$ are
\cite{dmkr} 
\ben 
A_1 = A_0 \cosh \alpha_1,~~T_1 = {T_0 / \cosh
\alpha_1}, ~~\phi_1 = \tanh \alpha_1.
\label{eq:znine}
\een
The emitted particle has energy $\omega_1$ and momentum $q_1$ along
$x^{11}$ 
\ben
\omega_1 = \omega_0 \cosh \alpha_1,~~~~~~~~
q_1 = \omega_0 \sinh \alpha_1,
\label{eq:zeleven}
\een
while the transverse momentum is unchanged (in Planck units).
In string theory this is a zero brane with quantized charge $Q_0=q_1 R_1$.
Note that the expression appearing in the thermal distribution function 
$\xi_0 = {\omega_0 / T_0} = {(\omega_1 - q_1 \phi_1)/ T_1} = \xi_1$
appears in the correct form. Using (\ref{eq:zthree}) and (\ref{eq:znine}) the 
absorption cross-section for the Hawking
particle is 
\ben
\sigma_1 = \sigma_0 {\omega_0 \over \omega_1}
= A_1 {\omega_1 - q_1 \phi_1 \over \omega_1}.
\label{eq:zfourteen}
\een

After the second transformation, the black hole is 
collection of nonextremal D4 branes of string theory 
along $(1234)$, or longitudinal
5-branes of M-theory along $(1234,11)$. Using standard T-duality formulae
the radius of $x^{11}$ in this M-theory becomes
\ben
R_2 = R_1 {l_1^4 \over L_1^{(1)}L_2^{(1)}L_3^{(1)}L_4^{(1)}}. 
\label{eq:newone}
\een
The area and the temperature
do not change, $A_2 = A_1, T_2 = T_1$.

The emitted particle is now an {\it extremal} 4-brane of string 
theory with some
transverse motion, energy $\omega_2 = \omega_1$ and the
quantized 4-brane charge is $Q_4 = Q_0$. To avoid explicit mention
of higher form gauge fields,
we will use an ``equivalent $x^{11}$ momentum'' for this 4-brane, defined 
as follows. We perform
T-dualities which convert this 4-brane into a 0-brane. In this case
these are T-dualities along $(1234)$. This results in a new underlying
M-theory with a $x^{11}$ radius
${\tilde R}_2$ which may be easily calculated using standard T-duality
formulae to yield
${\tilde R}_2 = R_1$. The emitted particle then has a $x^{11}$ momentum $q_2$
\ben
q_2 \equiv {Q_4 / {\tilde R}_2} = q_1.
\een
We will use this ``equivalent $x^{11}$ momentum'' in all the following steps
and denote it by $q_n$ and denote the corresponding M-theory 
radius by ${\tilde R}_n$.
Written in terms of the new quantities the thermal factor
is exactly what is expected, $\xi_2 = \xi_1 =  
(\omega_2 - q_2 \phi_1)/ T_2$.
The absorption cross-section is, of course unchanged, $\sigma_2 = \sigma_1$.

After the third step, the black hole is a D4 brane along $(1234)$ with 
some 0-brane charge,
or a nonextremal 5-brane with longitunal momentum in the language of 
M theory. Its area and temperature are
\ben
A_3 = A_2 \cosh \alpha_2,~~~~~T_3 = {T_2 / \cosh \alpha_2}.
\label{eq:zeighteen}
\een
The crucial point is that the nature of the emitted particle does not
change appreciably at this step. If the emitted four-brane was at rest
this would have been an extremal five brane in M-theory which is
invariant under boosts in the longitudinal direction. This means that
the metric produced by this object and its integer valued charge 
remains the same.
Since the radius of $x^{11}$ is Lorentz contracted,
total energy is decreased by the same factor
\ben
\omega_3 = {\omega_2 / \cosh \alpha_2}.
\label{eq:znineteen}
\een
Once again we need to find the equivalent $x^{11}$ momentum. One
finds that ${\tilde R}_3 = {\tilde R}_2 \cosh \alpha_2$ so that
\ben
q_3 = q_2 /\cosh \alpha_2.
\label{eq:ztwentyg}
\een
Essentially the same conclusion holds when there is a small transverse
momentum, as will be justified later. At this stage one has
$\xi_3 = \xi_2 = {(\omega_3 - q_3) \phi_1 / T_3}$.
Finally the cross
section is
\ben
\sigma_3 = \sigma_2 {\omega_2 \over \omega_3} = A_3 {\omega_3 - q_3 \phi_1
\over \omega_3},
\label{eq:twentyone}
\een
where we have used (\ref{eq:znineteen}) and (\ref{eq:ztwentyg}).

The remaining steps are repetitions of the above.
For the even-numbered T-duality steps, $\omega_{2n} = \omega_{2n-1},
q_{2n} = q_{2n-1}$
and $\sigma_{2n} = \sigma_{2n-1}$ while for the odd number steps
involving boosts $\omega_{2n+1} = \omega_{2n}/\cosh \alpha,
~~q_{2n+1} = q_n/\cosh \alpha$, where $\alpha$ is the relevant boost
parameter, while $\sigma_{2n-1}$ is related to
$\sigma_{2n-2}$ by the relation (\ref{eq:zthree}). 
Note that from step 4 onwards the T-dualities required to define
the equivalent $x^{11}$ momentum for the emitted particle are {\em not}
the same as the T-dualities in the previous step. Nevertheless, it is easy
to check that the above relations continue to hold.
The main point is that for low transverse momentum, the emitted particle 
carries only the charge which is imparted to it by the first boost
$\alpha_1$.

At the end of these steps we have a four dimensional black hole
(in the noncompact space $x^7 \cdots x^9$) made
of $D0$ branes along with three sets of intersecting D4 branes along
$(1234)$, $(1256)$ and $(3456)$ which is 
emitting D0 branes. The black hole is near-extremal when the boost
parameters $\alpha_n$ are large. 
In the language of M-theory we have three sets of five-branes
intersecting along $x^{11}$ and carrying some momentum along $x^{11}$,
emitting particles which carry momentum along $x^{11}$
equal to $q_8$ as well as a small transverse momentum. 
The black hole has
a four dimensional area 
\ben A_8 = A_0 \cosh \alpha_1 \cosh \alpha_2
\cosh \alpha_3 \cosh \alpha_4, 
\een 
and the energy and charge of the
emitted particle are 
\ben \omega_8 = {\omega_0 \cosh \alpha_1 \over
\prod_{i=2}^4\cosh \alpha_i},~~~~~~ q_8 = {\omega_0 \sinh \alpha_1
\over \prod_{i=2}^4\cosh \alpha_i}.
\label{eq:ztwentytwo}
\een
The cross-section obtained by the above procedure is
\ben
\sigma_8 = \sigma_0~{\omega_0 \omega_2 \omega_4 \omega_6
\over \omega_1 \omega_3 \omega_5 \omega_7} = 
A_8~{\omega_8 - q_8 \phi_1 \over \omega_8}.
\label{eq:ztwentythree}
\een
Note that after all these steps, $\phi_1$ has again become the potential
due to the rank-1 gauge field at the horizon to which the emitted 0-brane
couples. (\ref{eq:ztwentythree})
 is precisely the semiclassical answer for emission from the
four dimensional black hole with four charges \cite{gk}.

In \cite{gk} an effective microscopic model for the above emission
process has been proposed in the near extremal limit : this is a 
superconformal field theory with
$c = 6$  on a line parallel to the intersection of the branes, but with
 a length equal to the charges times the length of this direction. This
 model correctly
reproduces the semiclassical entropy.  Furthermore, the emission rates
for low energy scalars (in the noncompact four dimensional sense)
which may carry some momentum along $x^{11}$ was calculated 
in this microscopic theory along the
lines of \cite{callanmal} and the result is
exactly the semiclassical answer given by 
(\ref{eq:ztwentythree}) (with $\phi_1 = \tanh \alpha_1 \sim 1$,
since the hole is near-extremal).

We have related the low energy absorption cross section of
uncharged minimal scalars in the neutral hole to the low energy
charged scalar absorption for the four charge extremal hole. 
Thus the microscopic calculation of \cite{gk} also provides a
microscopic calculation of Hawking radiation from the four dimensional
Schwarzschild hole.

Note that the agreement in absorption properties
is not related to the fact that the cross section of
neutral quanta for {\it both } neutral and charged holes is the area
of the horizon and thus does not follow from the agreement of the 
entropies.  Rather, in our calculation, we observe first that
under the boost the neutral hole gets a charge, while at the same time
the cross section becomes {\it smaller} than the area.  This smaller
cross section is however just the right one to represent the emission
of the {\it charged} quanta from the charged hole. This pattern repeats
under the T-dualities/boosts, giving at the end a black hole carrying
four charges, and emitting particles carrying one of the four charges
(the momentum charge). The crucial point is that the various maps
which relate the neutral to the extremal hole also map, at the same
time, the emitted particle to something which can be treated easily in
the microscopic model.

It is important to note the restriction to low energies for the
emitted quantum. In fact we assume that $\omega\rightarrow 0$ is the
dominant limit, such that   $\omega
e^{\alpha_i}\rightarrow 0$, even though we must take
$\alpha_i$ large to approach the extremal
hole. Thus we do not obtain the greybody factors in this
calculation. The reason for this requirement on $\omega$ is the
following. At step 2 in the above sequence of calculations, we had a
4-brane that carried some transverse velocity. This is a 5-brane with
a transverse velocity in the 11-dimensional picture. What happens if
we now boost along $x^{11}$, as in step 3? If the transverse velocity
was zero, the 5-brane metric would remain unaffected, and the only
changes in the emitted quantum would come from the change of the scale
of recompactification. But if we have a transverse velocity of order
$\epsilon$, then the 5-brane that results after the boost is `tilted'
in the $x-x^{11}$ plane, by an angle $\sim \epsilon$ from the $x^{11}$
axis.  This is the picture in the covering space, but now we are
unable to do a Kaluza-Klein reduction along the $x^{11}$ direction,
since we have no translation invariance in the $x^{11 }$
direction. Thus we cannot do the recompactification. In the limit
$\epsilon\rightarrow 0$ we can ignore this `tilt', and do the
reduction, which is what we have implicitly done. To check the above
estimates, we first take the metric produced by a 5-brane with a small
transverse velocity - this
can be obtained from the Aichelburg-Sexl metric for a graviton,
followed by T-dualities. Then we boost in the $x^{11}$ direction, and
read off the `tilt' from the resulting solution.

We would like to thank S. Kalyana Rama for discussions. S.R.D. would like to
thank the Physics Department of Brown University and C.T.P., M.I.T. for
hospitality during the final stages of this work.
S.D.M. is supported
in part by D.O.E. cooperative agreement DE-FC02-94ER40818.

\end{document}